\documentclass[floatfix,twocolumn,aps,showpacs,showkeys,nofootinbib]{revtex4}

\usepackage{amssymb}
\usepackage{amsmath}
\usepackage{graphics}
\usepackage{epsfig}
\usepackage[dvips]{color}
\usepackage{bm}
\usepackage{color}

\newcommand{\be}{\begin{equation}}
\newcommand{\ee}{\end{equation}}
\newcommand{\bea}{\begin{eqnarray}}
\newcommand{\eea}{\end{eqnarray}}
\newcommand{\beas}{\begin{eqnarray*}}
\newcommand{\eeas}{\end{eqnarray*}}

\newcommand{\nn}{\nonumber\\}

\begin{document}
\title{Chiral transition with magnetic fields}
\author{Alejandro Ayala$^{1,4}$, Luis Alberto Hern\'andez$^{1,4}$, Ana J\'ulia Mizher$^1$, Juan Crist\'obal Rojas$^2$, Cristi\'an Villavicencio$^3$}
\affiliation{$^1$Instituto de Ciencias
  Nucleares, Universidad Nacional Aut\'onoma de M\'exico, Apartado
  Postal 70-543, M\'exico Distrito Federal 04510,
  Mexico.\\
  $^2$Departamento de F\'isica, Universidad Cat\'olica del Norte, Casilla 1280, Antofagasta, Chile\\
  $^3$Universidad Diego Portales, Casilla 298-V, Santiago, Chile.\\
  $^4$Centre for Theoretical and Mathematical Physics, and Department of Physics,
  University of Cape Town, Rondebosch 7700, South Africa}

\begin{abstract}

We study the nature of the chiral transition for an effective theory with spontaneous breaking of symmetry, where charged bosons and fermions are subject to the effects of a constant external magnetic field. The problem is studied in terms of the relative intensity of the magnetic field with respect to the mass and the temperature. When the former is the smallest of the scales, we present a suitable method to obtain magnetic and thermal corrections up to ring order at high temperature. By these means, we solve the problem of the instability in the boson sector for these theories, where the squared masses, taken as functions of the order parameter, can vanish and even become negative. The solution is found by considering the screening properties of the plasma, encoded in the resummation of the ring diagrams at high temperature. We also study the case where the magnetic field is the intermediate of the three scales and explore the nature of the chiral transition as we vary the field strength, the coupling constants and the number of fermions. We show that the critical temperature for the restoration of chiral symmetry monotonically increases from small to intermediate values of the magnetic fields and that this temperature is always above the critical temperature for the case when the magnetic field is absent. 

\end{abstract}



\maketitle

\section{Introduction}\label{I}

The nature of the QCD phase transitions has been a subject of great interest over the last decades. High quality data provided by the relativistic heavy-ion program carried out at the CERN Super Proton Synchrotron and under development at the BNL Relativistic Heavy Ion Collider and the CERN Large Hadron Collider has made it possible to test theoretical ideas about the properties of hadrons subject to extreme conditions of density and temperature. In addition, QCD on the lattice has produced results guiding, as well as complementing these findings with first principles calculations. Since observables are often defined in a regime where perturbative techniques offer little hope to describe the nature of strongly coupled systems, it has become important to resort to effective models to gain analytical insight about the QCD interaction within this environment.

More recently, there has been growing interest on the effects that a magnetic background may have on the QCD phase transitions. An external magnetic field can be viewed as a control parameter for the thermodynamics. Peripheral heavy-ion collisions can be used as a physically realizable situation to explore this possibility since they generate magnetic fields that are very intense during the very early stages of the collision, when they are estimated to be even stronger than those in magnetars. Magnetic fields are ubiquitous, appearing at all scales. They can influence the statistical properties of particles that make up these systems and can help catalyze the phase transitions in different contexts~\cite{Ayala1,Agasian:2008tb,Fraga:2008qn,Mizher:2008hf,Mizher:2010zb,Menezes:2008qt, Boomsma:2009yk,Fukushima:2010fe,Johnson:2008vna,Preis:2010cq,Callebaut:2011uc,Avancini:2011zz,Andersen:2011ip,Andersen:2012dz, Skokov:2011ib,Fraga:2012fs,Fukushima:2012xw,Ayala3,Simone}.

Lattice simulations have also paid attention to the QCD phase structure in the presence of magnetic fields. Early results showed that the critical temperature increased with the intensity of the magnetic field~\cite{Braguta, D'Elia:2011zu}. This result agreed with most of the model calculations. Later results, obtained by considering smaller lattice spacing and physical quark masses, found an opposite behavior~\cite{Fodor,Bali:2012zg}, which was afterwards also obtained in Ref.~\cite{Noronha} using the bag model and in Refs.~\cite{Farias, Ferreira1} postulating a magnetic field and temperature dependent running coupling in the Nambu-Jona-Lasinio model. The most recent results attribute such a decrease to a back reaction of the Polyakov loop, which indirectly feels the magnetic field and drives down the critical temperature for the chiral transition~\cite{Bruckmann:2013oba}. This kind of behavior has been obtained also in Ref.~\cite{Ferreira} using an extension of the Polyakov-NJL model.

An important ingredient in the study of the chiral transition is the development of a condensate as the system transits from the high to the low temperature phases. Within effective models, this condensate is described introducing boson degrees of freedom. When bosons are electrically charged their condensation is also subject to the influence of magnetic fields. The field theoretical treatment of the condensation of charged-boson systems at finite temperature in the presence of magnetic fields is plagued with subtleties. For example, it is well known that for massless bosons described at finite temperature, there is a divergence in the occupation number coming from the Bose-Einstein distribution at zero momentum. This divergence is suppressed by the momentum phase space factor in $d\geq 2$ spatial dimensions. In contrast, this divergence is enhanced in the presence of a magnetic field given that the energy levels separate into transverse and longitudinal (with respect to the magnetic field direction) and the former are accounted for in terms of discrete Landau levels. Therefore, the system experiences an effective dimensional reduction of the momentum integrals and thus the longitudinal mode alone is unable to tame the divergence of the Bose-Einstein distribution unless the system is described in a number of spatial dimensions $d>4$~\cite{May, Daicic, Elmfors}. The problem can be solved by a proper treatment of the physics involved when magnetic fields are introduced. For instance, it has recently been shown that even for $d=3$ it is possible to find the appropriate condensation conditions when accounting for the plasma screening effects~\cite{Ayala2, Loewe}.

Another subtlety that appears when magnetic fields are present occurs for systems whose condensate develops from an spontaneous breaking of symmetry. When the boson fields are expanded around the true minimum $v_0$, the squared of their mass, $m_b^2$, becomes a function of the order parameter $v$. $m_b^2$ can become negative for some values of $v$ in the domain range, $0\leq v\leq v_0$, which is of interest to describe the phase transition at finite temperature.  
When not properly treated, these negative values produce a non-analytic behavior of the vacuum energy. In a recent study~\cite{AHLMRV}, we have put forward a method to cure such misbehavior at high temperature for the case when the magnetic field strength is larger than the mass parameter of a purely boson theory. In this work we use a different approach suited for the case where the magnetic field strength is smaller or of the same order as the mass parameter. This is the infrared regime and the non-analyticity is cured by accounting for the plasma screening effects encoded in the contribution of the ring diagrams at high temperature. We account for the full ring dependence, as opposed to the case treated in Ref.~\cite{AHLMRV} where we carried out an expansion to first order in the self-energy to account for these contributions. This makes an important difference for the behavior of the critical temperature. We find that even for the purely boson sector the critical temperature is always above the corresponding critical temperature for zero magnetic field, whereas in Ref.~\cite{AHLMRV} we obtained that the critical temperature was below this critical temperature, albeit this function also increased with the field strength. We also include the effect of fermions and by these means explore the nature of the chiral phase transition as a function of the magnetic field strength, the couplings as well as the number of fermions. We show that the system presents first and second order phase transitions as we vary these parameters. If the system's phase transition is first order in the absence of the magnetic field, the latter produces that the transition eventually becomes second order as the field intensity increases. Given that the critical temperature is larger when the magnetic field is present, one can properly say that the development of the condensate is aided, that is to say, {\it catalyzed} by the field. 

The work is organized as follows: In Sec.~\ref{two} we introduce the model for the analysis, the so called Abelian Higgs model. We set up the calculation providing the general expressions for the one-loop effective potential, both for bosons and fermions. In Sec.~\ref{III} we explore in detail the effective potential in the weak field limit, namely, the case where the hierarchy of scales is $qB < |m_b|^2 < T^2$. We show that in this regime, it is necessary to account for the plasma screening effects which we include in terms of the so called ring diagrams. In Sec.~\ref{IV} we study the effective potential in the intermediate field case, where the hierarchy of scales is $ |m_b|^2 < qB < T^2$. In Sec.~\ref{V} we explore the parameter space looking for the values that produce either a first or second order phase transition. We also study the behavior of the critical temperature as a function of the magnetic field strength. We finally summarize and conclude in Sec.~\ref{concl}.

\section{One loop effective potential}\label{two}

To explore the interactions of charged bosons and fermions with an external magnetic field, we use as the effective tool the Abelian Higgs model with fermions. This model is given by the Lagrangian 
\bea
   {\mathcal{L}}&=&(D_{\mu}\phi)^{\dag}D^{\mu}\phi+i\bar{\psi}\gamma^\mu D_\mu\psi
   +\mu^{2}\phi^{\dag}\phi-\frac{\lambda}{4}   
   (\phi^{\dag}\phi)^{2}\nonumber\\
   &-&\frac{g}{\sqrt{2}}(\phi\bar{\psi}\psi + {\mbox{c.c.}}),
\label{lagrangian}
\eea
where $\phi$ and $\psi$ are charged scalar and fermion fields, respectively and
 \bea
   D_{\mu}=\partial_{\mu}+iqA_{\mu},
\label{dcovariant}
\eea
is the covariant derivative. $A^\mu$ is the vector potential corresponding to an external magnetic field directed along the $\hat{z}$ axis,
\bea
   A^\mu=\frac{B}{2}(0,-y,x,0),
\label{vecpot}
\eea
and $q$ is the particle's electric charge. $A^\mu$ satisfies the gauge condition $\partial_\mu A^\mu=0$. In the language of the covariant $R_\xi$ gauges, this gauge condition corresponds to $\xi=0$ and therefore the Goldstone mode does couple to the gauge field. Since the gauge field is taken as classical, we do not consider its fluctuations and thus no loops involving the gauge field in internal lines. In this sense for our purposes, the Abelian Higgs model serves as an effective tool to describe the thermodynamic properties of charged scalars and fermions in the presence of a constant magnetic field. The squared mass parameter $\mu^2$ and the self-couplings $\lambda$ and $g$ are taken to be positive.

We can write the complex field $\phi$ in terms of their real components $\sigma$ and $\chi$,
\bea
   \phi(x)&=&\frac{1}{\sqrt{2}}[\sigma(x)+i \chi(x)],\nn
   \phi^{\dag}(x)&=&\frac{1}{\sqrt{2}}[\sigma(x)-i\chi(x)].
\label{complexfield}
\eea
To allow for an spontaneous breaking of symmetry, we let the $\sigma$ field to develop a vacuum expectation value $v$
\bea
   \sigma \rightarrow \sigma + v,
\label{shift}
\eea
which can later be taken as the order parameter of the theory. After this shift, the Lagrangian can be rewritten as
\bea
   {\mathcal{L}} &=& -\frac{1}{2}[\sigma(\partial_{\mu}+iqA_{\mu})^{2}\sigma]-\frac{1}
   {2}\left(\frac{3\lambda v^{2}}{4}-\mu^{2} \right)\sigma^{2}-\frac{\lambda}{16}\sigma^{4}\nn
   &-&\frac{1}{2}[\chi(\partial_{\mu}+iqA_{\mu})^{2}\chi]-\frac{1}{2}\left(\frac{\lambda v^{2}}{4}-   
   \mu^{2} \right)\chi^{2}-\frac{\lambda}{16}\chi^{4} \nn
  &+&\frac{\mu^{2}}{2}v^{2}-\frac{\lambda}{16}v^{4}
  +i\bar{\psi}(\partial_{\mu}+iqA_{\mu})\psi-gv\bar{\psi}\psi+{\mathcal{L}}_{I},
  \label{lagranreal}
\eea
where ${\mathcal{L}}_{I}$ represents the Lagrangian describing the interactions among the fields $\sigma$, $\chi$ and $\psi$, after symmetry breaking. It is well known that for the Abelian Higgs model, with a local, spontaneously broken gauge symmetry, the gauge field $A^\mu$ acquires a finite mass and thus cannot represent the physical situation of a massless photon interacting with the charged scalar field. Therefore, for the discussion we ignore the mass generated for $A^\mu$ as well as issues regarding renormalization after symmetry breaking and concentrate on the scalar and fermion sectors. From Eq.~(\ref{lagranreal}) we see that the $\sigma$, $\chi$ and fermion masses are given by
\bea
  m^{2}_{\sigma}&=&\frac{3}{4}\lambda v^{2}-\mu^{2},\nn
  m^{2}_{\chi}&=&\frac{1}{4}\lambda v^{2}-\mu^{2}\nn
  m_f&=&gv.
  \label{masses}
\eea

\subsection{Tree plus one-loop effective potential}

The tree-level potential is given by
\be
  V^{{\mbox{\small{(tree)}}}}=-\frac{1}{2}\mu^{2}v^{2}+\frac{\lambda}{16}v^{4}.
\label{treelevel}
\ee
The minimum is obtained for
\be
  v_{0}=\frac{2\mu}{\sqrt{\lambda}}.
  \label{vefmin}
\ee
Notice that  
\bea
  \frac{d^{2}V^{(tree)}}{dv^{2}}&=&\frac{3\lambda v^{2}}{4}-\mu^{2}\nn
  &=&m_\sigma^2
  \label{condmass}
\eea
and also that the field $\chi$ corresponds to the Goldstone boson.

The expression for the one-loop effective potential for one boson field with squared mass $m_b^2$ at finite temperature $T$ in the presence of a constant magnetic field can be written as
\bea
  V_b^{(1)} &=& \frac{T}{2}\sum_n\int dm_b^2\int\frac{d^3k}{(2\pi)^3}\int_0^\infty
   \frac{ds}{\cosh (qBs)}\nn
   &\times&e^{-s(\omega_n^2+k_3^2 + k_\perp^2\frac{\tanh (qBs)}{qBs} + m_b^2)},
   \label{boson1}
\eea
where $\omega_n=2n\pi T$ are boson Matsubara frequencies. Performing the integration over $d^2k_\perp$, introducing the sum over Landau levels, integrating over $ds$, performing the sum over Matsubara frequencies and the integration over $dm_b^2$, we get
\bea
   V_b^{(1)}&=&\frac{1}{2}\frac{2qB}{4\pi}\sum_{l}\int\frac{dk_3}{2\pi}
   \left[ \omega_{l} + 2T\ln (1 - e^{-\omega_{l}/T})\right]\nn
   &\equiv&V_b^{(1,{\mbox{\tiny{vac}}})} + V_b^{(1,{\mbox{\tiny{matt}}})},
   \label{bosonV+T}
\eea
where
\bea
   \omega_{l}=\sqrt{k_3^2 + m_b^2 + 2(l + 1/2)qB}.
   \label{omegab}
\eea
Similarly, the expression for the one-loop effective potential for one fermion field with mass $m_f$ at finite temperature $T$ in the presence of a constant magnetic field can be written as
\bea
  V_f^{(1)}&=& -\sum_{r=\pm 1}T\sum_n\int dm_f^2\int\frac{d^3k}{(2\pi)^3}\int_0^\infty
   \frac{ds}{\cosh (qBs)}\nn
   &\times&e^{-s(\tilde{\omega}_n^2+k_3^2 + k_\perp^2\frac{\tanh (qBs)}{qBs} + m_f^2 +r qB)},
   \label{fermion1}
\eea
where $\tilde{\omega}_n=(2n+1)\pi T$ are fermion Matsubara frequencies. The sum over the index $r$ corresponds to the two possible spin orientations along the magnetic field direction. Performing the integration over $d^2k_\perp$, introducing the sum over Landau levels, integrating over $ds$, performing the sum over Matsubara frequencies and the integration over $dm_f^2$, we get
\bea
   V_f^{(1)}&=&-\frac{2qB}{4\pi}\sum_{l,r}\int\frac{dk_3}{2\pi}
   \left[ \omega_{lr} + 2T\ln (1 + e^{-\omega_{lr}/T})\right]\nn
   &\equiv&V_f^{(1,{\mbox{\tiny{vac}}})} + V_f^{(1,{\mbox{\tiny{matt}}})}
   \label{fermionV+T}
\eea
where
\bea
   \omega_{lr}=\sqrt{k_3^2 + m_f^2 + 2[l + (1 + r)/2]qB}.
   \label{omegaf}
\eea
Equations~(\ref{bosonV+T}) and~(\ref{fermionV+T}) are our master equations. On each of these, the first terms represent the magnetic-vacuum contributions whereas the second ones are the magnetic-matter contributions. We proceed to explore their behavior as we vary the strength of the magnetic field from small to intermediate values.

\section{Weak field limit}\label{III}

The magnetic-vacuum contributions, both for bosons and fermions, can be analytically expressed in closed form. This is shown in the appendix. For the present purposes and in order to explicitly show the cancellation of the infrared offending terms, let us first seek an approximation for the case where $qB < |m_b^2|$, $m_f^2$ with the temperature as the largest of the energy scales. 

We start with Eq.~(\ref{bosonV+T}). Note that we can write the sum over Landau levels as
\bea
   S_b\equiv\sum_l(2qB)g_l,
\label{almostEMboson}
\eea
where
\bea
   g_l\equiv \int\frac{dk_3}{2\pi}
   \left[ \omega_{l} + 2T\ln (1 - e^{-\omega_{l}/T})\right].
\label{gl}
\eea
From Eq.~(\ref{omegab}) we note that the increment in the summation index is $h=2qB$ and that the sum is evaluated at the midpoint between consecutive values of $l$. We can thus use the Euler-MacLauren approximation for the sum, written as
\bea
   S_b&=&\left\{
   \int dy\ g(y)-\frac{1}{2}
   \frac{B_2h^2}{2!}[g'(y=\infty) - g'(y=0)]
   \right\},\nn
\label{EMb}
\eea
where $B_2=1/6$ is the second Bernoulli number, $y=(2l+1)qB$ and we have kept the expression explicitly only up to ${\mathcal{O}}(h^2)$. Therefore, we can write
\bea
   V_b^{(1)}&=&\frac{m_b^4}{64\pi^2}
   \left[ \ln\left(\frac{m_b^2}{2\mu^2}\right) -\frac{1}{2}\right.\nn
   &+&\left. \ln\left(\frac{(4\pi T)^2}{m_b^2}\right)
   - 2\gamma_E + \frac{3}{2}\right]\nn
   &-&\frac{\pi^2T^4}{90} + \frac{m_b^2T^2}{24} - \frac{m_b^3T}{12\pi}\nn
   &-&\frac{(qB)^2}{192\pi^2}
   \left[ \ln\left(\frac{m_b^2}{2\mu^2}\right) + 1\right.\nn 
   &+&\ln\left(\frac{(4\pi T)^2}{m_b^2}\right)
   - 2\gamma_E -  \frac{2\pi T}{m_b}\nn
   &+&\left. \zeta (3)\left(\frac{m_b}{2\pi T}\right)^2 - \frac{3}{4}\zeta (5) \left(\frac{m_b}{2\pi T}\right)^4   
   \right],
\label{bosonB}
\eea
where $\gamma_E$ is the Euler gamma and $\zeta$ is the Riemann Zeta function. We have refrained from combining the terms coming from the vacuum and matter contributions to emphasize their origin. We have
also introduced a counter term $- \delta m_b^2 = m_b^2(1/\epsilon + \ln (2\pi) - \gamma_E)$ to take care of the boson mass renormalization and chosen the renormalization scale as $\tilde{\mu}=e^{-1/2}\mu$. The charge-field renormalization is explicitly performed in the appendix. Note that the terms proportional to odd powers of $m_b\equiv\sqrt{m_b^2}$ can potentially cause a non-analyticity when $m_b^2$ vanishes or becomes negative. As we will show, these terms are cancelled when considering the plasma screening properties encoded in the resummation of the {\it ring diagrams}.  

Let us now look at Eq.~(\ref{fermionV+T}). We can write the sum over Landau levels and the spin index as
\bea
   S_f\equiv\sum_{lr}(2qB)g_{lr}
\label{almostEMfermion}
\eea
where
\bea
   g_{lr}\equiv \int\frac{dk_3}{2\pi}
   \left[ \omega_{lr} + 2T\ln (1 + e^{-\omega_{lr}/T})\right].
\label{glr}
\eea
From Eq.~(\ref{omegaf}) the increment in the summation index is still $h=2qB$ but this time the sum is evaluated at the end points of consecutive values of $l$. We can thus use the Euler-MacLauren approximation for the sum, now written as
\bea
   S_f&=&\left\{
   \int dy\ \tilde{g}(y)+
   \frac{B_2h^2}{2!}[\tilde{g}'(y=\infty) - \tilde{g}'(y=0)]
   \right\},\nn
\label{EMf}
\eea
where $\tilde{g}(y)=2g(y)$, $y=2lqB$ and we have kept the expression explicitly only up to ${\mathcal{O}}(h^2)$. Therefore, we can write
\bea
   V_f^{(1)}&=&-\frac{m_f^4}{16\pi^2}\left[ \ln\left(\frac{m_f^2}{2\mu^2}\right) -
   \frac{1}{2}\right.
   \nn
   &+&\left.\ln\left(\frac{(\pi T)^2}{m_f^2}\right) - 2\gamma_E + \frac{3}{2}\right] - \frac{7\pi^2T^4}{180} + 
   \frac{m_f^2T^2}{12}\nn
   &-&\frac{(qB)^2}{24\pi^2}\left[ \ln\left(\frac{m_f^2}{2\mu^2}\right) + 1 +
   \ln\left(\frac{(\pi T)^2}{m_f^2}\right) - 2\gamma_E \right],\nn
\label{fermionB}
\eea
we have also  refrained from combining the terms coming from the vacuum and matter contributions to emphasize their origin. We have also introduced a counter term $- \delta m_f^2 = m_f^2(1/\epsilon + \ln (2\pi) - \gamma_E)$ to take care of the fermion mass renormalization. The charge-field renormalization is explicitly performed in the appendix.

\subsection{Ring diagrams}

The ring contribution to the effective potential represents the leading correction in the infrared for theories where there are massless boson modes. The dominant contribution is obtained from $\omega_n=0$. For a single boson field, this is given by
\bea
   V^{({\mbox{\small{ring}}})}_{b}=\frac{T}{2}
   \int \frac{d^3 k}{(2\pi)^3}\ln[1+\Pi\
   \Delta_B(\omega_n=0,k)],
   \label{ring}
\eea
where $ \Delta_B(\omega_n,k)$ is the Matsubara propagator in the presence a magnetic field. For the self-energy $\Pi$ we will consider the dominant contribution in the high temperature limit coming from the boson self-interaction as well as from its interaction with fermions~\cite{LeBellac}
\bea
   \Pi=\lambda\frac{T^2}{12}+N_fg^2\frac{T^2}{6},
   \label{selfenergy}
\eea
where we have allowed for an arbitrary number of fermions $N_f$. Note that the self-energy is momentum independent. We can write Eq.~(\ref{ring}) as
\bea
   V^{({\mbox{\small{ring}}})}_{b}=\frac{T}{2}
   \int \frac{d^3 k}{(2\pi)^3}\left[\ln(\Delta_B^{-1}+\Pi) - \ln(\Delta_B^{-1})\right].
   \label{ringexpanded}
\eea
The first (second) term in Eq.~(\ref{ringexpanded})  corresponds to taking the mass as $\Pi + m_b^2$ ($m_b^2$). We write $V^{({\mbox{\small{ring}}})}_{b}=V^{({\mbox{\small{ring}}})}_{b\ I}-V^{({\mbox{\small{ring}}})}_{b\ II}$. Let us compute explicitly the second term, 
\bea
   V^{({\mbox{\small{ring}}})}_{b\ II}&=&\frac{T}{2}
   \int \frac{d^3 k}{(2\pi)^3}\left[\ln(\Delta_B^{-1})\right]\nn
   &=&\frac{T}{2}
   \int \frac{d^3 k}{(2\pi)^3}\int dm_b^2 \frac{d}{dm_b^2}
   \left[\ln(\Delta_B^{-1})\right]\nn
   &=&\frac{T}{2}
   \int dm_b^2 \int \frac{d^3 k}{(2\pi)^3}
   \Delta_B.
   \label{ringexpandedII}
\eea
We can now proceed in the same manner as we did to go from Eq.~(\ref{boson1}) to Eq.~(\ref{bosonB}). The result is
\bea
   V^{({\mbox{\small{ring}}})}_{b\ II}&=&
   \left(\frac{T}{8\pi}\right)\left[ \frac{2}{3}(m_b^2 + \Lambda^2)^{3/2} - \frac{2}{3}m_b^3 + 
   \frac{(qB)^2}{12m_b}\right],\nn
   \label{II}
\eea
where for the ease of the computation we introduced the ultraviolet cutoff $\Lambda$. Therefore, we arrive at the expression for the ring contribution of a single boson species
\bea
   V^{({\mbox{\small{ring}}})}_{b}&=&V^{({\mbox{\small{ring}}})}_{b\ I}-V^{({\mbox{\small{ring}}})}_{b\ II}\nn
   &=&\left(\frac{T}{12\pi}\right)\left[ m_b^3 - (m_b^2 + \Pi)^{3/2}\right.\nn 
   &-& \left.
   \frac{(qB)^2}{8m_b} + \frac{(qB)^2}{8(m_b^2 + \Pi)^{1/2}}\right].
   \label{ringfin}
\eea
After adding up Eqs.~(\ref{bosonB}) and~(\ref{ringfin}), we confirm that the terms with odd powers of $m_b$ in Eq.~(\ref{bosonB}) cancel and in those, the boson mass is effectively substituted by the combination $(m_b^2 + \Pi)^{1/2}$. The self-energy encodes the plasma screening properties in the infrared region.

The final expression for the effective potential, considering the contribution from the $\sigma$, $\chi$ and $N_f$ fermion fields in the weak magnetic field regime is given by
\bea
   V^{({\mbox{\small{eff}}})}&=&
   -\frac{\mu^2}{2}v^2 + \frac{\lambda}{16}v^4\nn
   &+&\sum_{i=\sigma,\chi}\left\{\frac{m_i^4}{64\pi^2}\left[ \ln\left(\frac{(4\pi T)^2}{2\mu^2}\right) 
   -2\gamma_E +1\right]
   \right.\nn
   &-&\frac{\pi^2T^4}{90} + \frac{m_i^2T^2}{24} - \frac{(m_i^2+\Pi)^{3/2}T}{12\pi}\nn
   &-&\frac{(qB)^2}{192\pi^2}\left[ \ln\left(\frac{(4\pi T)^2}{2\mu^2}\right)
   - 2\gamma_E + 1\right.\nn
   &-&\frac{2\pi T}{(m_i^2+\Pi)^{1/2}}\nn
   &+&\left.\left. 
   \zeta (3)\left(\frac{m_b}{2\pi T}\right)^2 - \frac{3}{4}\zeta (5) \left(\frac{m_b}{2\pi T}\right)^4   
   \right]\right\}\nn
   &-&N_f\left\{\frac{m_f^4}{16\pi^2}\left[ \ln\left(\frac{(\pi T)^2}{2\mu^2}\right)
   -2\gamma_E +1\right]\right.\nn
   &+&\frac{7\pi^2T^4}{180} - \frac{m_f^2T^2}{12}\nn
   &+&\left.\frac{(qB)^2}{24\pi^2}\left[ \ln\left(\frac{(\pi T)^2}{2\mu^2}\right)
   - 2\gamma_E + 1\right]\right\},
   \label{Veff}
\eea
where $m_\sigma^2$, $m_\chi^2$, $m_f$ and $\Pi$ are given by Eqs.~(\ref{masses}) and~(\ref{selfenergy}), respectively. Note that Eq.~(\ref{Veff}) is free from non-analiticities when $m_\sigma^2$, $m_\chi^2$ become zero or even negative. When $v=0$, the combination $m_b^2+\Pi$ ($b=\sigma$, $\chi$) $\rightarrow -\mu^2 + \Pi$. Therefore for $\Pi$ given by Eq.~(\ref{selfenergy}), we see that Eq.~(\ref{Veff}) is valid provided that 
\bea
   T > \frac{\mu}{\sqrt{\frac{\lambda}{12} + \frac{N_fg^2}{6}}}.
\label{validity}
\eea

\section{Intermediate field regime}\label{IV}

For the regime where $|m_b^2|$, $m_f^2 \sim qB < T^2$ we apply the lessons learned during the analysis of the weak field case where we found that the infrared offending terms come exclusively from the Matsubara boson mode with $n=0$. This means that when looking for an approximation one needs to treat this mode separate from the others. 

\subsection{Bosons}

We start from Eq.~(\ref{boson1}) written in the form
\bea
   V^{(1)}_b&=&\frac{T}{2}\sum_n\int\frac{d^3k}{(2\pi)^3}
   \ln [\Delta_B(\omega_n, k)^{-1}],\nn
   &=&
   \frac{T}{2}
   \sum_n\int dm_b^2\int\frac{d^3k}{(2\pi)^3}
   \Delta_B (\omega_n, k).
\label{finiteT1}
\eea 
For the $n\neq 0$ modes in Eq.~(\ref{finiteT1}) one can resort to expanding the Matsubara propagator in powers of $qB/ T^2$, in the same fashion as was done  in Ref.~\cite{Ayala4}. Nevertheless, for $n=0$, use of this approximation would amount to describing the situation where $qB \ll |m^2_i|$. To avoid such limitation, we treat the zero frequency separately. In this way
\begin{eqnarray}
   \sum_n\Delta_B(\omega_n, k) =
   \sum_{n\neq 0}\Delta_B(\omega_{n\neq 0}, k)+
   \Delta_B(0, k).
   \label{modeseparation}
\end{eqnarray}
For the first term on the right-hand side of Eq.~(\ref{modeseparation}) we use  the  expansion as in Ref.~\cite{Ayala4}, valid for the case where $qB$, $m_b^2\ll T^2$ but that otherwise, when excluding the zero mode, does not assume a hierarchy between $m^2_b$ and $qB$,
\begin{eqnarray}
\Delta_B(\omega_{n\neq 0}, k) &\approx& 
   \frac{1}{\omega_n^2+k^2+m_b^2}\nn
   &\times&\left[ 1 - \frac{(qB)^2}
   {(\omega_n^2+k^2+m_b^2)^2}\right.\nn
   &+&
   \left.\frac{2(qB)^2k_\perp^2}{(\omega_n^2+k^2+m_b^2)^3}\right].
   \label{nonzeromode}
\end{eqnarray}
For the second one we keep the Schwinger proper time expression in Eucledian space for $n=0$, that is
\begin{eqnarray}
\Delta_B(0, k)=
 \int_{0}^{\infty} \frac{ds}{\cos (qBs)}e^{-is\left[k_{z}^2 + k_\perp^2\frac{\tan
   (qBs)}{qBs} + m_b^2 -i\epsilon\right]}.\nn
\label{zeromode}
\end{eqnarray}
Inserting Eqs.~(\ref{zeromode}) and~(\ref{nonzeromode}) into Eq.~(\ref{finiteT1}) we obtain
\bea
   V_{b}^{(1)}=V_{b\ I}^{(1)} + V_{b\ II}^{(1)}
\label{finiteT2}
\eea
where
\bea
  V_{b\ I}^{(1)}&\equiv&\frac{T}{2}\sum_{n\neq 0}\int dm_i^2\int \frac{d^3 k}{(2\pi)^3}
  \Delta_{B}(\omega_{n\neq 0},k),
  \label{nonzero_mode_pot}
\eea
and
\bea
  V_{b\ II}^{(1)}&\equiv&\frac{T}{2}\int dm_b^2 \int \frac{d^3 k}{(2\pi)^3}
  \Delta_{B}(\omega_{n=0},k).
  \label{zero_mode_pot}
\eea

\subsection{Non-zero modes}

We can explicitly carry out the sum and integrals in Eq.~(\ref{nonzero_mode_pot}). The sum over the non-zero modes is performed by means of the Mellin technique~\cite{Bedingham}. Under these conditions, after mass renormalization and up to ${\mathcal{O}}(m_b^4)$ we get
\begin{eqnarray}
   V_{b\ I}^{(1)}&=&
   \frac{m_b^2T^2}{24} +
   \frac{m_b^4}{64\pi^2}\left[\ln\left(\frac{(4\pi T)^2}{2\mu^2}\right)  -2\gamma_E +  1\right]\nn
   &-& \frac{(qB)^2}{192\pi^2}\left[\zeta (3)\left(\frac{m_b}{2\pi T}\right)^2
   - \frac{3}{4}\zeta (5)\left(\frac{m_b}{2\pi T}\right)^4\right],\nn 
\label{omegaT}
\end{eqnarray}
where we choose the renormalization scale, like before, as $\tilde{\mu}=e^{-1/2}\mu$. Equation~(\ref{omegaT}) coincides with Eq.~(\ref{bosonB}), except for $v$-independent terms, which are obtained upon integrating over $m_b^2$ after including the integration constant.

\subsection{Zero mode plus ring}

We start by noticing that the ring contribution, Eq.~(\ref{ring}), can be written as
\bea
   V^{({\mbox{\small{ring}}})}_{b}&=&\frac{T}{2}\int dm_b^2
   \int \frac{d^3 k}{(2\pi)^3}\frac{d}{dm_b^2}\nn
   &\times&\Big\{\ln[ \Delta_B(\omega_n=0,k)^{-1} + \Pi]\nn
   &-& \ln[ \Delta_B(\omega_n=0,k)^{-1}]\Big\}\nn
   &=&\frac{T}{2}\int dm_b^2
   \int \frac{d^3 k}{(2\pi)^3}\left[\Delta_B(\omega_n=0,k)^{-1} + \Pi \right]^{-1}\nn
   &-&\frac{T}{2}\int dm_b^2
   \int \frac{d^3 k}{(2\pi)^3}\Delta_B(\omega_n=0,k),
   \label{ringStrong}
\eea
where $\Pi$ is the boson self-energy, given by Eq.~(\ref{selfenergy}). Note that the second term in Eq.~(\ref{ringStrong}), cancels the zero-mode contribution, Eq.~(\ref{zero_mode_pot}). Furthermore, since the combination $\left[\Delta_B(\omega_n=0,k)^{-1} + \Pi \right]^{-1}$ corresponds to evaluating the propagator with the substitution $m_b^2\rightarrow m_b^2 + \Pi$, then the combined contributions from the zero-mode and the ring diagrams can be expressed in terms of the propagator
\bea
   \Delta_B=\int_0^\infty
   \frac{ds}{\cosh (qBs)}
   e^{-s(\omega_n^2+k_3^2 + k_\perp^2\frac{\tanh (qBs)}{qBs} + m_b^2 + \Pi)}.\nn
\label{interms}
\eea
Performing the momentum integrals, the integral over the mass parameter and introducing the Landau levels, we obtain for the contribution of the zero mode plus ring
\begin{eqnarray}
   V_{b\ II}^{(1)}+V^{({\mbox{\small{ring}}})}_{b}=
   \frac{T(2qB)^{3/2}}{8\pi^2}\zeta\left(-\frac{1}{2},\frac{1}{2}+\frac{m_b^2+\Pi}{2qB}\right),\nn
\label{omegaTB}
\end{eqnarray}
where $\zeta(s,q)$ is the Hurwitz Zeta function.

\subsection{Fermions}

We now use an approximation to write the fermion propagator suited for the case $qB \sim m_f^2$. For this purpose we use the findings in Ref.~\cite{Chyi} valid in the {\it weak field limit}, that is, where $qB\ll T^2$ but that otherwise does not assume a hierarchy between $qB$ and $m_f^2$. The fermion contribution to the effective potential up to one-loop can then be written as
\bea
   V_f^{(1)}&=&-2T\sum_n\int dm_f^2\int\frac{d^3k}{(2\pi)^3}
   \left\{
   \frac{1}{\tilde{\omega}_n^2+k^2+m_f^2}\right.\nn 
   &+& \left.\frac{2(qB)^2k_\perp^2}{(\tilde{\omega}_n^2+k^2+m_f^2)^4}
   \right\}.
\label{oneloopstrongfer}
\eea
Performing the sum over the Matsubara frequencies, the integration over the fermion mass and over the momentum we obtain in the high temperature approximation and after mass and charge renormalization
\bea
   V_f^{(1)}&=&-\frac{m_f^4}{16\pi^2}\left[ \ln\left(\frac{(\pi T)^2}{2\tilde{\mu}^2}\right) 
   - 2\gamma_E  \frac{}{}\right]\nn
   &-&\frac{7\pi^2T^4}{180} + \frac{m_f^2T^2}{12}\nn
   &-&\frac{(qB)^2}{24\pi^2}\left[ \ln\left(\frac{(\pi T)^2}{2\tilde{\mu}^2}\right)
   - 2\gamma_E \right].
\label{fermionlargeB}
\eea
Note that Eq.~(\ref{fermionlargeB}) coincides with Eq.~(\ref{fermionB}). This could come as a surprise given that Eq.~(\ref{fermionB}) was explicitly obtained within the approximation $qB < m_f^2$ whereas Eq.~(\ref{fermionlargeB}) was obtained without reference to a hierarchy between $qB$ and $m_f^2$ and only under the assumption that $T^2\gg m_f^2,\ qB$. A closer look to the derivation of these equations reveals the reason for them to coincide. First, unlike the boson case, the fermion contribution does not have a zero mode and thus there is no need to treat this mode separately. Second, the term $\ln[(\pi T)^2/\tilde{\mu}^2]$ appearing in both the $qB$-independent and dependent terms comes from two different contributions, the vacuum one and the matter one and can be written as
\bea
   \ln\left(\frac{(\pi T)^2}{\tilde{\mu}^2}\right) &=& {\mbox{vac}} + {\mbox{matt}},\nn
   {\mbox{vac}} &=& \ln\left(\frac{m_f^2}{\tilde{\mu}^2}\right),\nn
   {\mbox{matt}} &=& \ln\left(\frac{(\pi T)^2}{m_f^2}\right).
   \label{log}
\eea
For the vacuum contribution, the analysis requires establishing the hierarchy between the only two relevant scales $m_f^2 > qB$, whereas for the matter contribution at high $T$, the calculation requires $T^2\gg m_f^2,\ qB$. When combining the vacuum and matter contributions, all reference to the mass scale in this logarithmic term disappears, effectively making that the only surviving comparison of scales is between $T^2$ and $qB$, and that the calculation be valid for $T^2\gg m_f^2,\ qB$. In fact, note that in this regime, the leading contribution from the magnetic field in either of Eqs.~(\ref{fermionB}) or Eq.~(\ref{fermionlargeB}) is independent of $m_f$, which makes more evident that to obtain this contribution there is no need to establish a hierarchy between $qB$ and $m_f^2$.

Including the $v$-independent terms and choosing the renormalization scale as $\tilde{\mu}=e^{-1/2}\mu$, the effective potential in the intermediate field regime can be written as
\bea
   V^{({\mbox{\small{eff}}})}&=&
   -\frac{\mu^2}{2}v^2 + \frac{\lambda}{16}v^4\nn
   &+&\sum_{i=\sigma,\chi}\left\{\frac{m_i^4}{64\pi^2}\left[ \ln\left(\frac{(4\pi T)^2}{2\mu^2}\right) 
   -2\gamma_E +1\right]
   \right.\nn
   &-&\frac{\pi^2T^4}{90} + \frac{m_i^2T^2}{24}\nn
   &+&\frac{T(2qB)^{3/2}}{8\pi}\zeta\left(-\frac{1}{2},\frac{1}{2}+\frac{m_i^2+\Pi}{2qB}\right)\nn
   &-&\frac{(qB)^2}{192\pi^2}\left[ \ln\left(\frac{(4\pi T)^2}{2\mu^2}\right)
   - 2\gamma_E + 1\right.\nn
   &+&\left.\left. 
   \zeta (3)\left(\frac{m_b}{2\pi T}\right)^2 - \frac{3}{4}\zeta (5) \left(\frac{m_b}{2\pi T}\right)^4   
   \right]\right\}\nn
   &-&N_f\left\{\frac{m_f^4}{16\pi^2}\left[ \ln\left(\frac{(\pi T)^2}{2\mu^2}\right)
   -2\gamma_E +1\right]\right.\nn
   &+&\frac{7\pi^2T^4}{180} - \frac{m_f^2T^2}{12}\nn
   &+&\left.\frac{(qB)^2}{24\pi^2}\left[ \ln\left(\frac{(\pi T)^2}{2\mu^2}\right)
   - 2\gamma_E + 1\right]\right\}.
   \label{Veff-mid}
\eea
Note that Eq.~(\ref{Veff-mid}) coincides with Eq.~(\ref{Veff}) after the replacement
\bea
   &-& \frac{(m_b^2+\Pi)^{3/2}T}{12\pi} -\frac{(qB)^2}{96\pi^2}\frac{\pi T}{(m_b^2+\Pi)^{1/2}}
   \rightarrow\nn
   &&\frac{T(2qB)^{3/2}}{8\pi}\zeta\left(-\frac{1}{2},\frac{1}{2}+\frac{m_b^2+\Pi}{2qB}\right)\nn
\label{replacement}
\eea
For the Hurwitz zeta function $\zeta(-1/2,z)$ in Eq.~(\ref{Veff-mid}) to be real, we need that 
\bea
   -\mu^2 + \Pi > qB,
\label{requirealso}
\eea
condition that comes from requiring that the second argument of the Hurwitz zeta function satisfies $z>0$, even for the lowest value of $m_b^2$ which is obtained for $v=0$. Furthermore, for the large $T$ expansion to be valid, we also require that
\bea
   qB/T^2 <1.
\label{otherrequirement}
\eea
Notice that in case we were to study the large field regime, namely, the case where $|m_b^2|$, $m_f^2 < T^2 \lesssim qB$, one can use the exact expressions for the magnetic-vacuum contribution in Eqs.~(\ref{bosonV+T}) and~(\ref{fermionV+T}) which, as we show in the appendix, can be written as
\bea
   V_b^{(1,{\mbox{\tiny{vac}}})}&=&-\frac{m_b^4}{64\pi^2}\ln\left(\frac{\mu^2}{qB}\right)
   + \frac{(qB)^2}{192\pi^2}\ln \left(\frac{\mu^2}{qB}\right)\nn
   &+& \frac{(qB)^2}{8\pi^2}\zeta'\left(-1,\frac{1}{2}+\frac{m_b^2}{2qB}\right),
\label{bosmagvac}
\eea
and
\bea
   V_f^{(1,{\mbox{\tiny{vac}}})}&=&\frac{(qB)m_f^2}{8\pi^2}\ln\left(\frac{m_f^2}{2qB}\right)
   +\frac{m_f^4}{16\pi^2}\ln\left(\frac{\mu^2}{qB}\right)\nn
   &+& \frac{(qB)^2}{24\pi^2}\ln\left(\frac{\mu^2}{qB}\right)\nn
   &-& \frac{(qB)^2}{2\pi^2}\zeta'\left(-1,1+\frac{m_f^2}{2qB}\right),
\label{fermagvac}
\eea
where $\zeta'(-1,x)=\left(\partial /\partial y\right)\zeta(y,x)|_{y=-1}$. We also show in the appendix that the weak field limit of Eqs.~(\ref{bosmagvac}) and~(\ref{fermagvac}) coincide with the corresponding terms in Eqs.~(\ref{bosonB}) and~(\ref{fermionB}), respectively. The magnetic-matter contributions to Eqs.~(\ref{bosonV+T}) and~(\ref{fermionV+T}) however cannot be easily approximated since, although $T$ is large, a large temperature expansion is not applicable. The reason is that, as the index $l$ becomes large, the quantity $lqB$ eventually becomes larger than $T^2$. One needs to resort instead to numerically find these last contributions.  Note that care has to be taken to also separate the boson Matsubara zero mode and replace it for the ring corrected contribution, in order to consistently follow the evolution of observables such as the critical temperature as functions of the magnetic field strength. Nevertheless, since in this work we have in mind to look at situations where the magnetic field is not the largest of the scales, we do not explore in detail this regime.

\begin{figure}[t!]
\begin{center}
\includegraphics[scale=0.6]{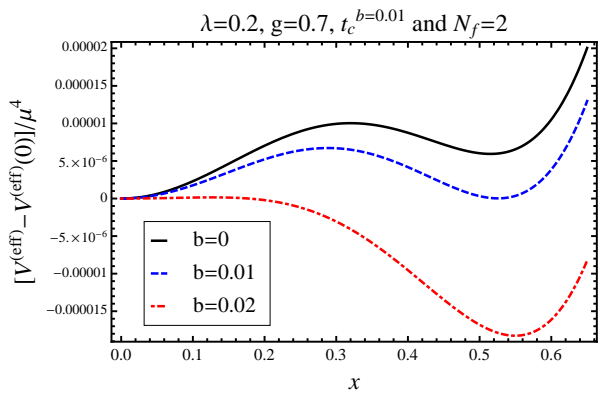}
\end{center}
\caption{Color on-line. Effective potential as a function of $x=v/\mu$ for $b=qB/\mu^2=0,\ 0.01,\ 0.02$ for $t_c^{b=0.01}=T_c^{b=0.01}/\mu$ and fixed values of $\lambda=0.2$ and $g=0.7$ and $N_f=2$. For the chosen set of parameters, the phase transition is first order.}
\label{fig1}
\end{figure}

\section{Parameter space}\label{V}

\begin{figure}[b!]
\begin{center}
\includegraphics[scale=0.6]{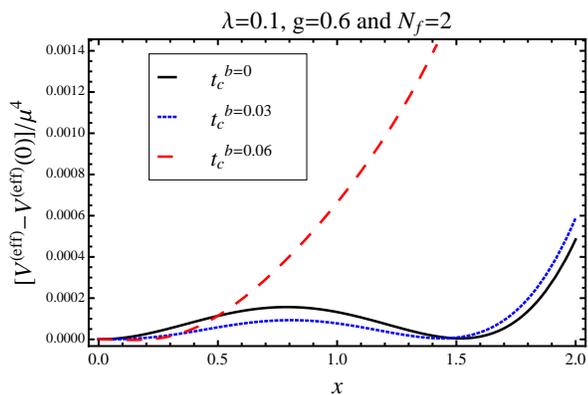}
\end{center}
\caption{Color on-line. Effective potential as a function of $x=v/\mu$ for $b=qB/\mu^2=0,\ 0.03,\ 0.06$ for fixed values of $\lambda=0.1$ and $g=0.6$ and $N_f=2$. Starting from first order, the phase transition becomes second order as we increase the field strength.}
\label{fig2}
\end{figure}

We now proceed to vary the parameters in the effective potential to explore the phase structure. Figure~\ref{fig1} shows the effective potential as a function of $v$ in units of $\mu$ ($x=v/\mu$), in the weak field limit, Eq.~(\ref{Veff}), for three values of $qB$ in units of $\mu^2$, ($b=qB/\mu^2$), computed for a given temperature $t_c^{b=0.01}=T_c^{b=0.01}/\mu$, which is the critical temperature (in units of $\mu$) when $b=0.01$, and fixed values of $\lambda=0.2$ and $g=0.7$, with $N_f=2$. We note that the figure shows a first order phase transition since at $t_c^{b=0.01}$ there are two minima separated by a barrier. We also note that a finite value of $b$ aids the phase transition since the system does not need to wait until the temperature is lowered to transit from the symmetric to the broken symmetry phase, as in the case with $b=0$. This point is further emphasized by looking at the curve with $b=0.02$ where we note that the phase transition already happened at a larger temperature. Therefore one speaks of a phase transition catalyzed by the magnetic field. 

\begin{figure}[t!]
\begin{center}
\includegraphics[scale=0.6]{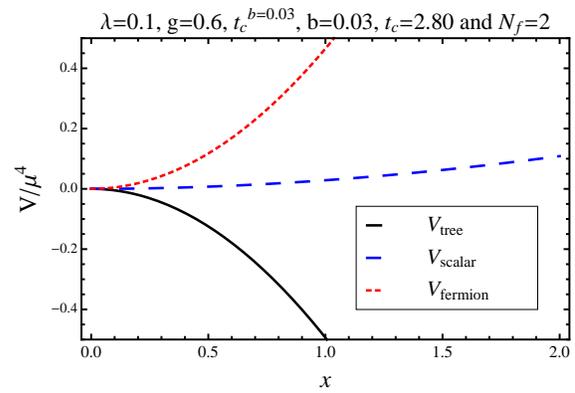}
\end{center}
\caption{Color on-line. Separate contributions from the tree-level, the boson one-loop (ring corrected) and fermion one-loop to the effective potential, for $\lambda=0.1$, $g=0.6$, $b=0.03$, $N_f=2$ evaluated at the critical temperature $t_c=2.8$. For these values, the fermion contribution overcomes the boson's and the combined effect is to produce a small hump that signals a first order phase transition.}
\label{fig3}
\end{figure}

\begin{figure}[b]
\begin{center}
\includegraphics[scale=0.6]{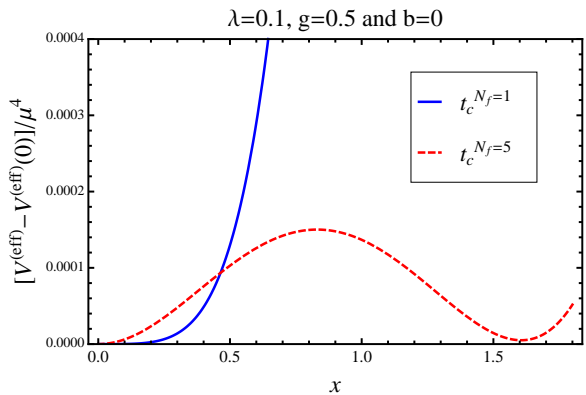}
\end{center}
\caption{Color on-line. Effect of the number of fermions on the order of the phase transition for $\lambda=0.1$, $g=0.5$ and $b=0$. If the phase transition is second order for $N_f=1$ this becomes first order as we increase the number of fermions to $N_f=5$.}
\label{fig4}
\end{figure}

The first order nature of the phase transition is due to the fermions and happens starting from $b=0$ when $g>\lambda$. However, the transition becomes second order as we increase the field strength. This is shown in fig.~\ref{fig2} for another choice of parameters also obeying $g>\lambda$, $\lambda=0.1$, $g=0.6$ and the same number of fermions $N_f=2$. In this figure the curves are computed each at its critical temperature. The phase transition for the curve with the largest value of the magnetic field  is second order  whereas the transition is first order for smaller values of the magnetic field.

That the contribution of fermions is responsible for the first order nature of the transition, for given combinations of the parameters, is illustrated in fig.~\ref{fig3}. This figure shows the separate contributions from the tree-level, the boson one-loop (ring corrected) and fermion one-loop to the effective potential, computed for $\lambda=0.1$, $g=0.6$, $b=0.03$, $N_f=2$ evaluated at the critical temperature $t_c=2.8$. Note that in the absence of fermions, the boson contribution pushes the effective potential toward negative values and that its strength is larger than the tree-level contribution. However, the fermion contribution pushes the effective potential toward positive values with an even larger strength to overcome the boson contribution and therefore the combined effect is to produce a small hump that signals the first order nature of the phase transition. Since the fermion contribution is proportional to $N_f$ and to $g$, a first order phase transition is more likely when these parameters grow.

\begin{widetext}
\begin{figure*}[t!]
\begin{center}
\begin{tabular}{ccc}
\includegraphics[scale=0.43]{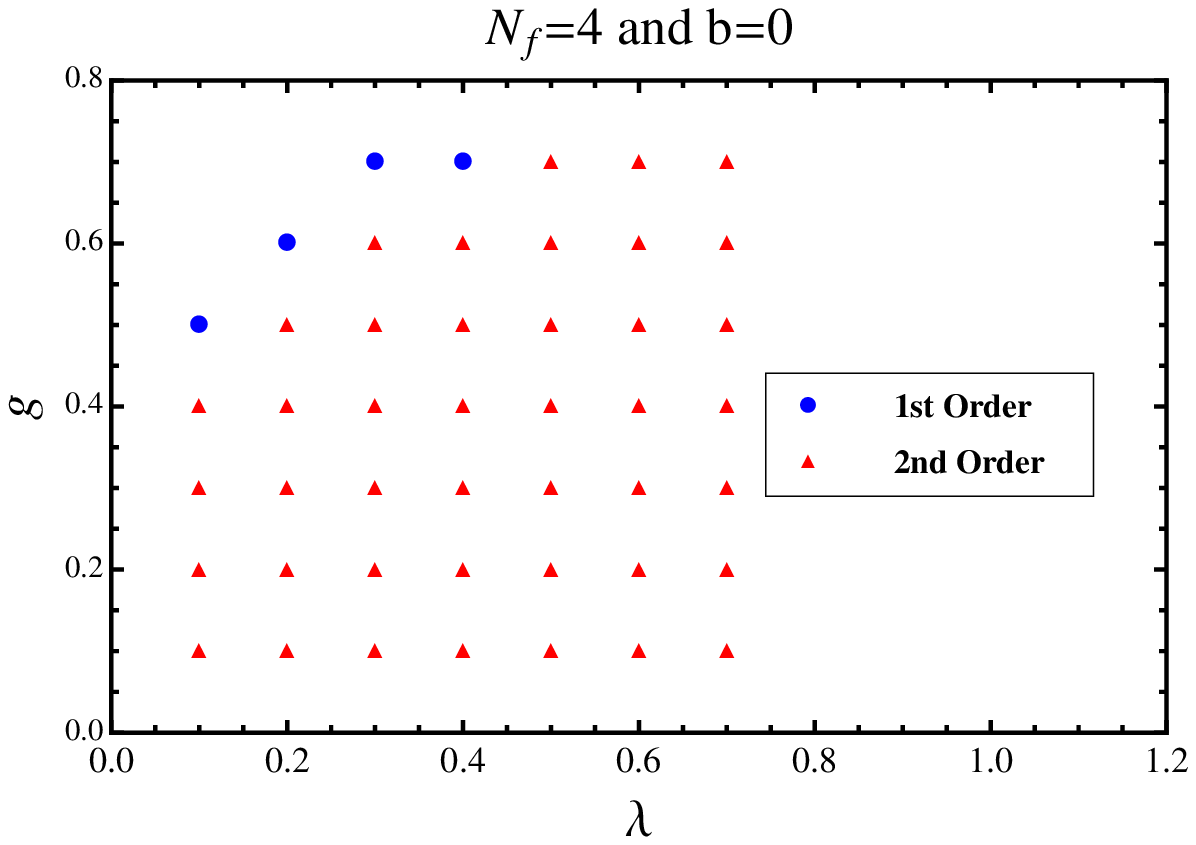} &
\includegraphics[scale=0.43]{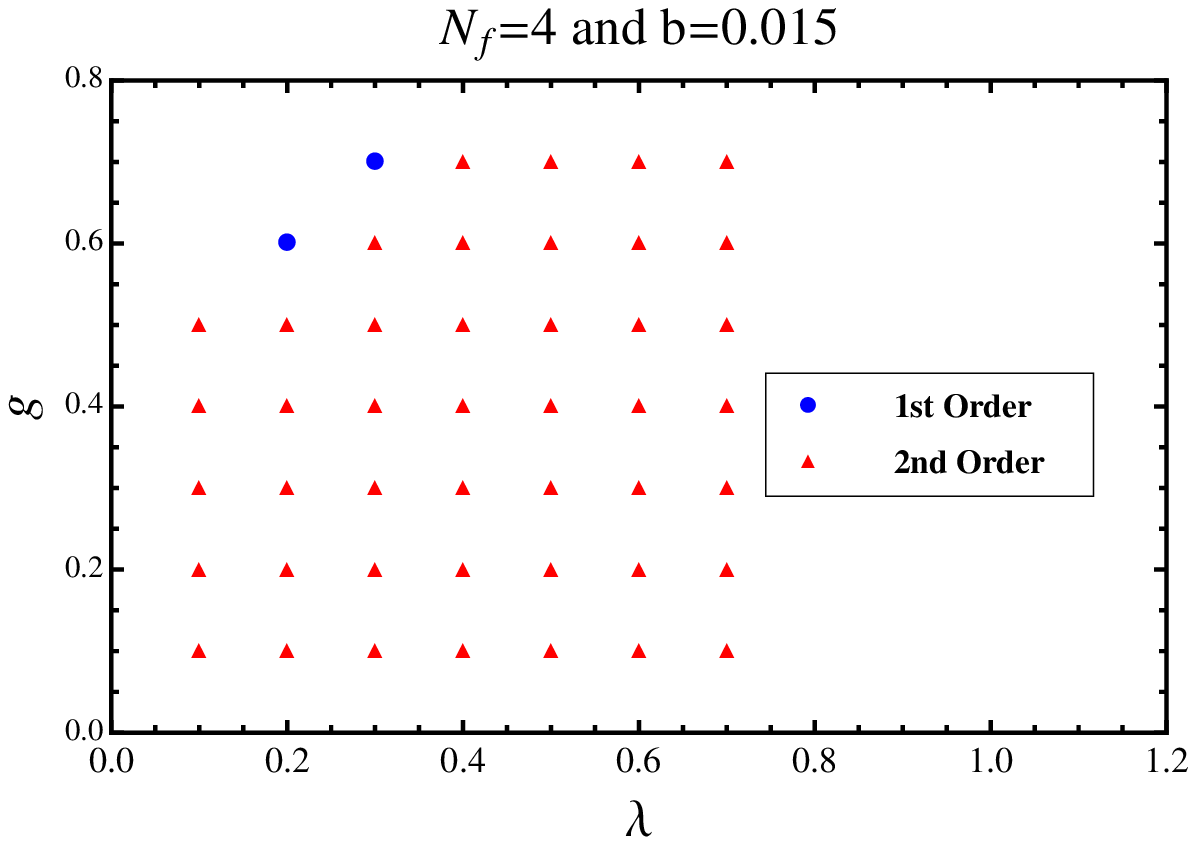} &
\includegraphics[scale=0.43]{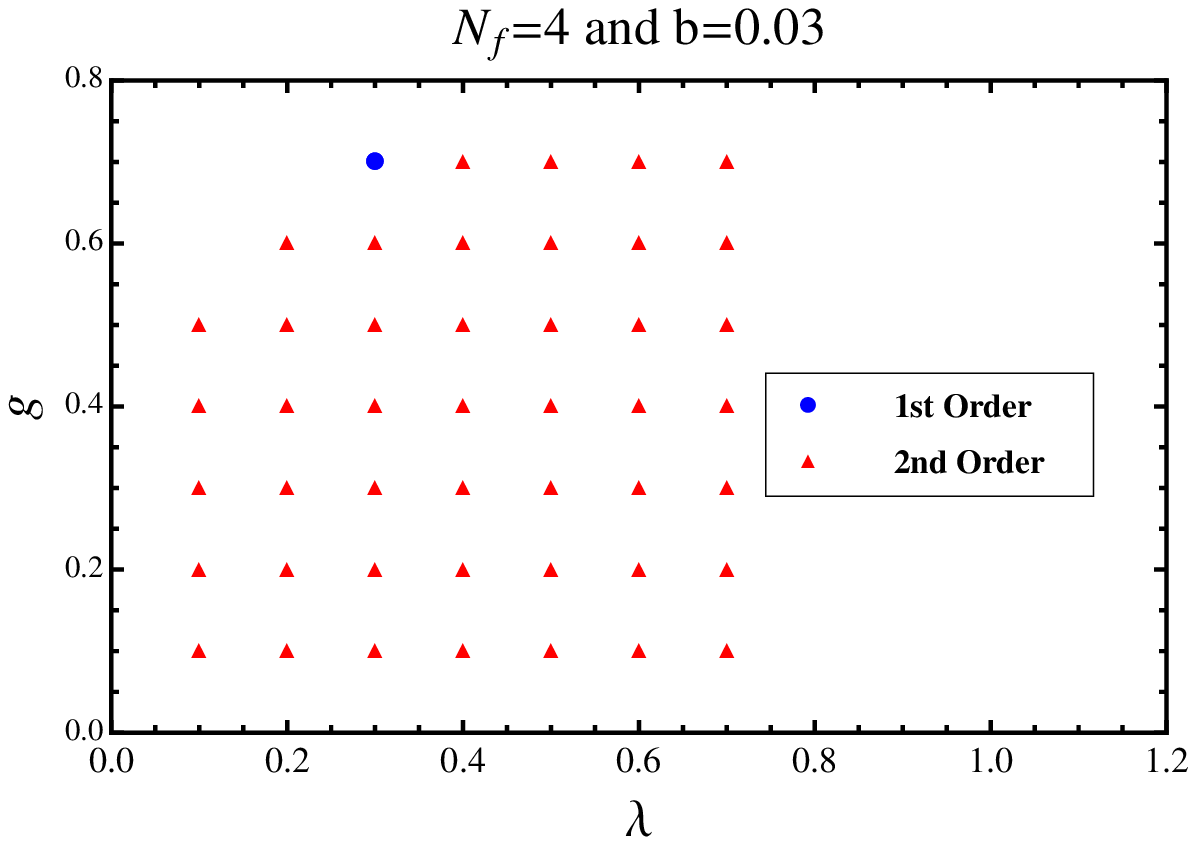}
\\ (a) &(b) &(c)
\end{tabular}
\end{center}
\caption{Color on-line. Phase diagram as a function of $\lambda$ and $g$ for different values of $b$: (a) $b=0$, (b) $b=0.015$ and (c) $b=0.03$, with $N_f=4$. As the field strength grows, the corner of the parameter space that allows a first order phase transition disappears.}
\label{fig5}
\end{figure*}
\end{widetext}

Figure~\ref{fig4} shows the effect of the number of fermions on the order of the phase transition for $\lambda=0.1$, $g=0.5$ and $b=0$. If the phase transition is second order for a small $N_f(=1)$ this becomes first order as we increase $N_f(=5)$. We emphasize that even if we start with a first order phase transition, the transition becomes second order as we increase the magnetic field strength. The above findings are summarized in fig.~\ref{fig5} where we show the phase diagram as we vary the parameters $\lambda$ and $g$ for different values of $b$ with a fixed number $N_f=4$. Note that as the field strength grows, the corner of the parameter space that allows a first order phase transition disappears.

\begin{widetext}
\begin{figure*}[t!]
\begin{center}
\begin{tabular}{ccc}
\includegraphics[scale=0.42]{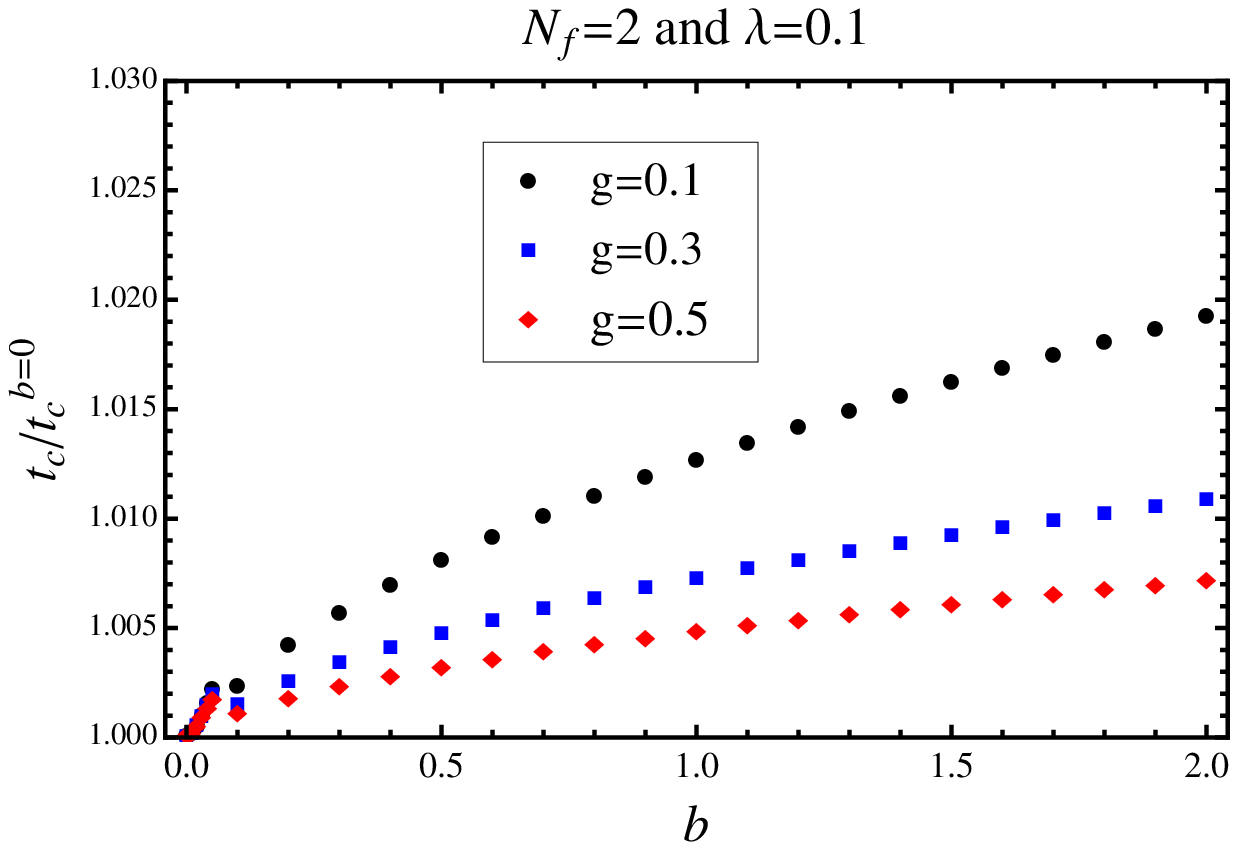} &
\includegraphics[scale=0.42]{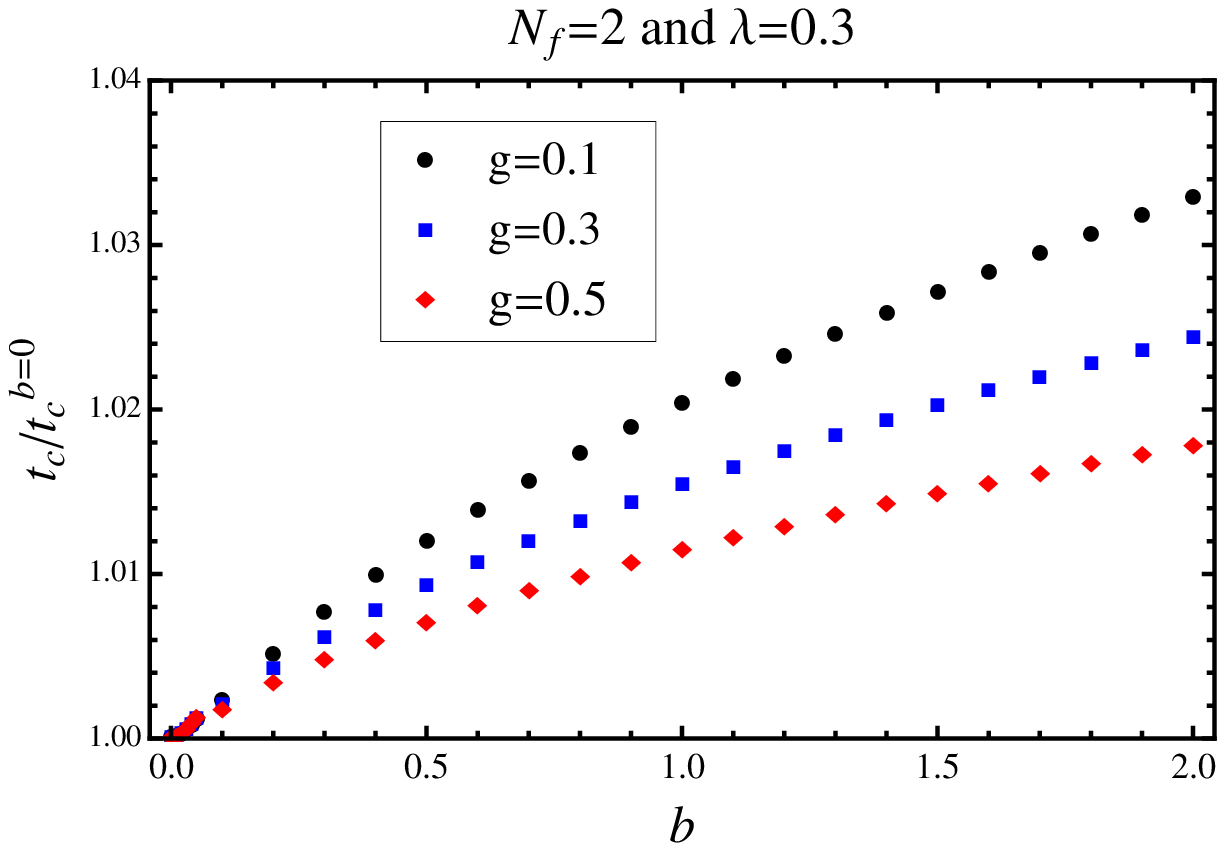} &
\includegraphics[scale=0.42]{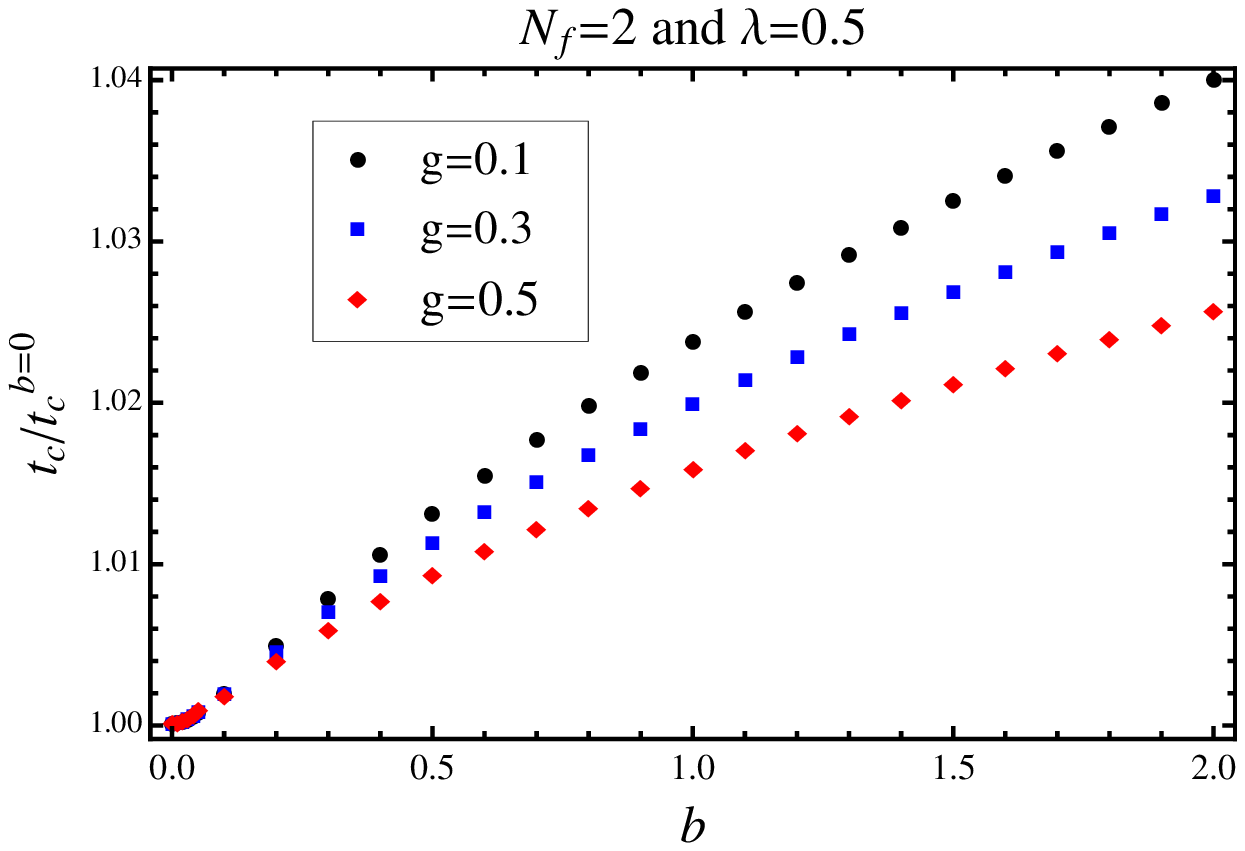}
\\ (a) &(b) &(c)
\end{tabular}
\end{center}
\caption{Color on-line. Critical temperature $t_c$ as a function of the field strength $b$ in the weak, $b\leq 0.05$, and intermediate, $0.05 < b \leq 2$, field regimes for $N_f=2$, three values of $g=0.1\ ,0.3\ ,0.5$, and (a) $\lambda=0.1$, (b) $\lambda=0.3$ and (c)  $\lambda=0.5$. The critical temperature is always a monotonically increasing function of the field strength. The curves join more smoothly as the couplings grow.}
\label{fig6}
\end{figure*}
\end{widetext}

Figure~\ref{fig6}$a$ shows the critical temperature $t_c$ as a function of the field strength $b$ in the weak, $b\leq 0.05$, and intermediate, $0.05 < b \leq 2$, field regimes for $N_f=2$, $\lambda=0.1$ and three values of $g=0.1\ ,0.3\ ,0.5$. The critical temperature is obtained computing the temperature for degenerate minima of the effective potential when the phase transition is first order and form setting the second derivative of the effective potential to zero at $v=0$ when the phase transition is second order. Note that the critical temperature is an monotonically increasing function of $b$ for all values of $g$. However, for this case, that corresponds to a low value of $\lambda$, the transition from the weak to the intermediate field regimes is not smooth. The reason is that the Euler-Maclaurin approximation that accounts for the magnetic field contribution for $b\simeq 0$ worsens as the field strength increases, given that the dominant contribution is proportional to $(m_b^2 + \Pi)^{-1/2}$, according to Eq.~(\ref{Veff}), and this factor grows close to $t_c$ since $m_b^2 + \Pi$ can become small for small values of $\lambda$ and $g$. The effect is magnified by increasing the value of the magnetic field. The situation improves if the coupling constants grow. This is shown in figs.~\ref{fig6}$b$ and~\ref{fig6}$c$ where we show $t_c$ as a function of the field strength in the weak and intermediate field regimes for $N_f=2$, the same three values of $g=0.1\ ,0.3\ ,0.5$ and $\lambda=0.3,\ 0.5$, respectively. The curves computed from the weak field limit and the intermediate field regime join smoothly for the cases considered.

\section{Summary and conclusions}\label{concl}

In this work we have studied the chiral phase transition at finite temperature for a system consisting of charged fermions and scalars with spontaneous breaking of symmetry and subject to the effects of a uniform magnetic field. For the analysis we have computed the finite temperature effective potential at one-loop, taking care of the non-analyticities in the infrared region by the resummation of the boson ring diagram contributions. By these means we have shown that, although the boson squared mass can become zero or even negative, the system is well behaved when accounting for the plasma screening properties encoded in the boson self-energy. We have studied the cases where the magnetic field satisfies the hierarchy of scales $qB < |m_b^2|  < T^2$ (weak field limit) and $|m_b^2| < qB < T^2$ (intermediate field regime) and the system's phase structure as we vary the couplings, the number of fermions and the strength of the magnetic field. We have shown that the system presents first and second order phase transitions depending on the values of the parameters. The first order nature of the phase transition is caused by the fermion contribution and happens when the coupling constant $g$ between fermions and bosons is larger than the boson self-coupling $\lambda$. This result is different from our previous findings~\cite{AHLMRV} where for the purely boson case we obtained first order phase transitions for small values of $\lambda$ and small field intensities. We can trace back this behavior to our previous poor approximation of the ring diagram contribution which in the present study has been corrected, accounting for the full ring dependence, as opposed to the case treated in Ref.~\cite{AHLMRV} where we carried out an expansion to first order in the self energy to account for these contributions. When the phase transition is first order for zero magnetic field, the latter makes the transition turn into second order. If the phase transition is second order for zero field, it continues being second order in the presence of the field. The critical temperature is a monotonic increasing function of the field strength. This is in contrast to recent lattice results~\cite{Fodor, Bali:2012zg} that find the critical temperature for chiral symmetry restoration to be a decreasing function of the field strength for large fields. Since the critical temperature increases with increasing magnetic field strength, our results show that the field catalyses the development of the condensate. This result is also generically obtained in non-perturbative treatments, such as Schwinger-Dyson techniques, of the QCD quark condensate~\cite{AB}. In this sense, the lattice results in Refs.~\cite{Fodor, Bali:2012zg} seem to go against the magnetic catalysis phenomenon obtained in many contexts. 

In conclusion, we have shown that the phase diagram for a system of charged fermions and scalars in the presence of a magnetic field has a rich structure. Our proper handling of the case where $qB<|m_b^2|$ provides a powerful result that ensures that even for small field strengths the critical temperature for chiral symmetry restoration increases and that this behavior continues for larger field strengths. The result makes it interesting to keep on looking for the origin of the lattice findings regarding the behavior of the critical temperature as a function of the field intensity. This will be the subject of a forthcoming work.

\section*{Acknowledgments}

The authors are in debt to R. Zamora for a careful reading of the manuscript as well as for very useful comments. Support for this work has been received in part from DGAPA-UNAM under grant number PAPIIT-IN103811, CONACyT-M\'exico under grant number 128534 and FONDECYT under grant  numbers 1130056 and 1120770. 

\section*{Appendix}

To compute the magnetic-vacuum contribution to Eq.~(\ref{bosonV+T}) we start from the expression for the one-loop vacuum piece for a charged boson in terms of Schwinger's proper-time method in Eucledian space
\bea
  V_b^{(1,{\mbox{\tiny{vac}}})} &=& \frac{1}{2}\int dm_b^2\int\frac{d^4k}{(2\pi)^4}\int_0^\infty
   \frac{ds}{\cosh (qBs)}\nn
   &\times&e^{-s(k_0^2+k_3^2 + k_\perp^2\frac{\tanh (qBs)}{qBs} + m_b^2)}.
   \label{bosonvacuum}
\eea
The momentum integrals are computed using dimensional regularization in $d\rightarrow 4$ dimensions. Introducing the renormalization scale $\tilde{\mu}$ and after carrying out the integration over the transverse components, we get
\bea
   V_b^{(1,{\mbox{\tiny{vac}}})} &=& \frac{qB}{(4\pi)}\tilde{\mu}^{4-d}\int dm_b^2\int\frac{d^{d-2}k_{\parallel}}{(2\pi)^{(d-2)}}
   \nn
   &\times&\sum_{l=0}\int_0^\infty ds e^{-s(k_{\parallel}^2 + (2l+1)qB + m_b^2)}
\label{aftertransv}
\eea
where we have introduced the sum over Landau levels. Performing the integration over $dm_b^2$ and over $d^{d-2}k_{\parallel}$ and after a change of variable $s\rightarrow t=s[(2l+1)qB + m_b^2]$, we get
\bea
   V_b^{(1,{\mbox{\tiny{vac}}})} &=& -\frac{qB}{(4\pi)^2}\frac{\tilde{\mu}^{4-d}}{(4\pi)^{d/2-2}}
   \int_0^\infty dt\ t^{-d/2}e^{-t}\nn
   &\times&\sum_{l=0}\frac{1}{[(2l+1)qB + m_b^2]^{1-d/2}}\nn
   &=&-\frac{2(qB)^2}{(4\pi)^2}\left(\frac{4\pi\tilde{\mu}^2}{2qB}\right)^{2-d/2}\nn
   &\times&\Gamma(1-d/2)\zeta\left(1-d/2,\frac{1}{2}+\frac{m_b^2}{2qB}\right).
\label{afterm}
\eea
Taking $d\rightarrow 4-2\epsilon$
\bea
   V_b^{(1,{\mbox{\tiny{vac}}})}&=&-\frac{2(qB)^2}{(4\pi)^2}\left(\frac{4\pi\tilde{\mu}^2}{2qB}\right)^{\epsilon}\nn
   &\times&\Gamma(-1+\epsilon)\zeta\left(-1+\epsilon,\frac{1}{2}+\frac{m_b^2}{2qB}\right).
\label{afterd}
\eea
Using that 
\bea
   \Gamma(-1+\epsilon)&\simeq& \frac{1}{\epsilon}-1+\gamma_E,\nn
   \zeta\left(-1+\epsilon,\frac{1}{2}+\frac{m_b^2}{2qB}\right)&\simeq&
   \zeta\left(-1,\frac{1}{2}+\frac{m_b^2}{2qB}\right)\nn
   &+& \epsilon\ \zeta'\left(-1,\frac{1}{2}+\frac{m_b^2}{2qB}\right),
\label{identities}
\eea
we get
\bea
   V_b^{(1,{\mbox{\tiny{vac}}})} &=& \frac{(qB)^2}{8\pi^2}
   \left\{
   \left[
   \frac{1}{\epsilon} + \ln (2\pi) - \gamma_E + 1 +\ln\left(\frac{\tilde{\mu}^2}{qB}\right)
   \right]\right.\nn
   &\times&\left. \zeta\left(-1,\frac{1}{2}+\frac{m_b^2}{2qB}\right)+ \zeta'\left(-1,\frac{1}{2}+\frac{m_b^2}{2qB}\right)
   \right\}.\nn
\label{afteridentities}
\eea
Introducing a counter term $-\delta qB^2=(qB)^2(1/\epsilon+ \ln (2\pi) - \gamma_E)$ to take care of charge-field renormalization and taking the renormalization scale as $\tilde{\mu}=e^{-1/2}\mu$, we get
\bea
   V_b^{(1,{\mbox{\tiny{vac}}})} &=& \frac{(qB)^2}{8\pi^2}
   \left\{
   \ln\left(\frac{\mu^2}{qB}\right)
   \zeta\left(-1,\frac{1}{2}+\frac{m_b^2}{2qB}\right)\right.\nn
   &+& \left.\zeta'\left(-1,\frac{1}{2}+\frac{m_b^2}{2qB}\right)
   \right\}.
\label{afterrenorm}
\eea
We now make use of the identity
\bea
   \zeta(-1,a)=-\left(\frac{1}{2}\right)\left(a^2 - a + \frac{1}{6}\right),
\label{zetaid}
\eea
to finally write Eq.~(\ref{afterrenorm}) as
\bea
   V_b^{(1,{\mbox{\tiny{vac}}})} &=& -\frac{m_b^4}{64\pi^2}\ln\left(\frac{\mu^2}{qB}\right)
   + \frac{(qB)^2}{192\pi^2}\ln\left(\frac{\mu^2}{qB}\right)\nn
   &+& \frac{(qB)^2}{8\pi^2}\zeta'\left(-1,\frac{1}{2}+\frac{m_b^2}{2qB}\right).
\label{bosbvacfin}
\eea
Equation~(\ref{bosbvacfin}) is valid for any value of $qB$. It is interesting however to take the weak field limit $qB/m_b^2\rightarrow 0$ to check whether this coincides with the magnetic vacuum contribution in Eq.~(\ref{bosonB}). For these purposes we use the asymptotic expansion~\cite{Elizalde}
\bea
   \zeta'(-1,u)\simeq\frac{u^2}{2}\ln u - \frac{1}{4}u^2- \frac{u}{2}\ln u + \frac{1}{12}\ln u + \frac{1}{12},
   \label{expansion}
\eea
valid for large $u$, together with
\bea
   \ln \left(\frac{1}{2}+\frac{1}{2x}\right)\simeq \ln \left(\frac{1}{2x}\right) + x - \frac{x^2}{2},
\label{together}
\eea
valid for small $x$, to arrive at
\bea
      V_b^{(1,{\mbox{\tiny{vac}}})} &\stackrel{\frac{qB}{m_b^2}\rightarrow 0}{\longrightarrow}& 
      \frac{m_b^4}{64\pi^2}\left[\ln\left(\frac{m_b^2}{2\mu^2}\right)-\frac{1}{2}\right]\nn
      &-&\frac{(qB)^2}{192\pi^2}\left[\ln\left(\frac{m_b^2}{2\mu^2}\right)+1\right].
\label{finalsmallB}
\eea
Equation~(\ref{finalsmallB}) coincides with the contribution from one boson species to the magnetic-vacuum in Eq.~(\ref{bosonB}).


The magnetic-vacuum contribution to Eq.~(\ref{fermionV+T}) follows steps similar to the boson case. We also start from the expression for the one-loop vacuum piece for a charged fermion in terms of Schwinger's proper-time method in Eucledian space
\bea
  V_f^{(1,{\mbox{\tiny{vac}}})} &=& -2\int dm_f^2\int\frac{d^4k}{(2\pi)^4}\int_0^\infty
   ds\nn
   &\times&e^{-s(k_0^2+k_3^2 + k_\perp^2\frac{\tanh (qBs)}{qBs} + m_f^2)}.
   \label{fermionvacuum}
\eea
The momentum integrals are computed using dimensional regularization in $d\rightarrow 4$ dimensions. Introducing the renormalization scale $\tilde{\mu}$ and after carrying out the integration over the momentum components, we get
\bea
   V_f^{(1,{\mbox{\tiny{vac}}})} &=& -\frac{2qB}{(4\pi)^{d/2}}\tilde{\mu}^{4-d}\int dm_f^2
   \int_0^\infty ds\nn
   &\times&\left[e^{-sm_f^2}+2\sum_{l=0} e^{-s[m_f^2+2(l+1)qB}\right]
\label{fermionaftertransv}
\eea
where we have introduced the sum over Landau levels. Performing the integration over $dm_f^2$ and over $ds$, we get
\bea
   V_f^{(1,{\mbox{\tiny{vac}}})} &=& \frac{2qB}{(4\pi)^{d/2}}\tilde{\mu}^{4-d}
   \Gamma(1-d/2)\Big\{(m_f^2)^{d/2-1}\nn
   &+&2\sum_{l=0}\left[m_f^2 + 2(l+1)qB\right]^{d/2-1}\Big\}.\nn
   &=& \frac{2qB}{(4\pi)^2}\Gamma(1-d/2)\left\{m_f^2\left(\frac{m_f^2}{4\pi\tilde{\mu}^2}\right)^{d/2-2}\right.\nn
   &+&\left.4qB\left(\frac{2qB}{4\pi\tilde{\mu}^2}\right)^{d/2-2}\zeta\left(1-d/2,1+\frac{m_f^2}{2qB}
   \right)\right\}.\nn
   \label{fermionafterm}
\eea
Taking $d\rightarrow 4-2\epsilon$
\bea
   V_f^{(1,{\mbox{\tiny{vac}}})}&=& \frac{2qB}{(4\pi)^2}\Gamma(-1+\epsilon)\left\{m_f^2\left(\frac{m_f^2}{4\pi\tilde{\mu}^2}\right)^{\epsilon}\right.\nn
   &+&\left.4qB\left(\frac{2qB}{4\pi\tilde{\mu}^2}\right)^{\epsilon}\zeta\left(-1+\epsilon,1+\frac{m_f^2}{2qB}
   \right)\right\}.\nn
\label{fermionafterd}
\eea
In the limit $\epsilon\rightarrow 0$ and using Eqs.~(\ref{identities}), we get
\bea
   V_f^{(1,{\mbox{\tiny{vac}}})} &=& \frac{2qB}{(4\pi)^2}
   \left\{m_f^2
   \left[
   -\frac{1}{\epsilon} - \ln (2\pi) + \gamma_E\right.\right.\nn
   &-&\left. 1 -\ln\left(\frac{2\tilde{\mu}^2}{m_f^2}\right)
   \right]\nn
   &+& 4qB\left[\left(
   -\frac{1}{\epsilon} - \ln (2\pi) + \gamma_E\right.\right.\nn
   &-&\left.1 -\ln\left(\frac{2\tilde{\mu}^2}{m_f^2}\right)\right)
   \zeta\left(-1,1+\frac{m_f^2}{2qB}\right)\nn
   &-&\left.\left.\zeta'\left(-1,1+\frac{m_f^2}{2qB}\right)\right]
   \right\}.\nn
\label{fermionafteridentities}
\eea
Introducing counter terms $-\delta qB^2=(qB)^2(1/\epsilon+ \ln (2\pi) - \gamma_E)$ and $-\delta m_f^2=m_f^2(1/\epsilon+ \ln (2\pi) - \gamma_E)$ to take care of charge-field and fermion mass renormalization and taking the renormalization scale as $\tilde{\mu}=e^{-1/2}\mu$, we get
\bea
   V_f^{(1,{\mbox{\tiny{vac}}})} &=& -\frac{(qB)}{8\pi^2}m_f^2\ln\left(\frac{2\mu^2}{m_f^2}\right)\nn
   &-&\frac{(qB)^2}{2\pi^2}\left[\ln\left(\frac{\mu^2}{qB}\right)
   \zeta\left(-1,1+\frac{m_f^2}{2qB}\right)\right.\nn
   &-&\left. \zeta'\left(-1,1+\frac{m_f^2}{2qB}\right)\right]
\label{fermionafterrenorm}
\eea
Once again we make use of the identity in Eq.~(\ref{zetaid}) to finally write Eq.~(\ref{fermionafterrenorm}) as
\bea
   V_f^{(1,{\mbox{\tiny{vac}}})} &=& \frac{(qB)}{8\pi^2}m_f^2\ln\left(\frac{m_f^2}{2qB}\right) + 
   \frac{m_f^4}{16\pi^2}\ln\left(\frac{\mu^2}{qB}\right)\nn 
   &+& \frac{(qB)^2}{24\pi^2}\ln\left(\frac{\mu^2}{qB}\right)\nn
   &-&  \frac{(qB)^2}{2\pi^2}\zeta'\left(-1,1+\frac{m_f^2}{2qB}\right)
\label{ferbvacfin}
\eea
Equation~(\ref{ferbvacfin}) is valid for any value of $qB$. It is interesting however to take the weak field limit $qB/m_f^2\rightarrow 0$ to check whether this coincides with the magnetic vacuum contribution in Eq.~(\ref{fermionB}). Using Eqs.~(\ref{expansion}) and~(\ref{together}) we arrive at
\bea
      V_f^{(1,{\mbox{\tiny{vac}}})} &\stackrel{\frac{qB}{m_f^2}\rightarrow 0}{\longrightarrow}& 
      -\frac{m_f^4}{16\pi^2}\left[\ln\left(\frac{m_f^2}{2\mu^2}\right)-\frac{1}{2}\right]\nn
      &-&\frac{(qB)^2}{24\pi^2}\left[\ln\left(\frac{m_f^2}{2\mu^2}\right)+1\right].
\label{finalsmallB}
\eea
Equation~(\ref{finalsmallB}) coincides with the contribution from one fermion species to the magnetic-vacuum in Eq.~(\ref{fermionB}).


\end{document}